*Entering new markets-a challenge in times of crisis*

**ANCA GHEORGHU**
anca.gheorghiu@gmail.com
**ANDA GHEORGHIU**
anda.gheorghiu@gmail.com

Hyperion University

**Abstract**

*After September 2008, the advanced economies severe decline caused demand for emerging economies' exports to drop and the crisis became truly global, much deeper and broader than expected.*

*In these times of global depression, most countries and companies are affected, some more than others. The financial crisis has turned out to be much deeper and broader than expected.*

*Entering new markets has always been a hazardous entrepreneurial attempt, but also a rewarding one, in the case of success. The paper aims to asses the market entry risk of a company trying to make a good acquisition, to buy shares of another company, activating in a foreign country. For this purpose, the case of Electroputere S.A., the old Romanian producer of railway equipment, has been chosen. The data were collected from the records of Bucharest Stock Exchange. After two years from the acquisition, one can draw a conclusion whether the strategy of the investor was a good one or a waste of money.*

**1. Preamble**

The financial crisis which started in 2007 and is still ongoing in 2010 has been considered the worst financial crisis since the Great Depression of the 1930s, since it affected drastically most countries in the world, big corporations and contributed to declines in consumer wealth and a noteworthy downward spiral in economic activity.

After September 2008, the advanced economies severe decline caused demand for emerging economies' exports to drop and the crisis became truly global, much deeper and broader than expected.

In these times of global depression, most countries and companies are affected, some more than others. The financial crisis has turned out to be much deeper and broader than expected.

But the crises do not bring out only losers; there are opportunities for savvy companies able to involve the entire staff in supporting new market entry strategies, guidelines for increased efficiency, by minimizing risks and cutting costs in key areas.

The attempt to enter new markets has always been a risky entrepreneurial act, but also a rewarding one, in the case of success. Producing profits from selling or producing abroad are great achievements for firms trying to expand their market share in foreign countries. However, companies should think very thoroughly their market entry strategies, in order to avoid the negative impact of some risks like: country risk, contractual risks, currency risk, environmental risks, etc.

**2. Description of a tool used to asses the prospective of success or failure for a company trying to enter a new market**

An useful tool to asses the prospective of success or failure for a company trying to enter a new market by using an associative strategy is I, a risk indicator defined[1] as follows:

$$I = N \times F \times \frac{1+RI_T}{1+RI_O} \times \frac{1+RCE_T}{1+RCE_O} \times \frac{V_{\text{company X}}}{CS_{\text{company X}}}$$

where the factors are:
- -the country-risk rating of the target-market (N)
- -the degree of cultural and organizational compatibility (F)
- -the inflation rate for the target-market ( $RI_T$ )
- -the inflation rate for the country of origin ( $RI_O$ )
- -the rate of economic growth for the target-country ( $RCE_T$ )
- -the rate of economic growth for the country of origin ( $RCE_O$ )
- -the social capital of the X firm, the patrimonial entity that is used for settling an associative strategy ($CS_{\text{X firm}}$)
- -the economic value of the enterprise ($V_{\text{X firm}}$).

If the society is listed at the stock exchange , than $V_{\text{X firm}}$ is the market value.

The rating N assess the risk degree of the target-country; it can take values between 1-10 (10 for stabile countries and with a good quality of the credit and 1 for the economic crisis situation and incertitude concerning payments and disbursements).

---

[1] *Gheorghiu Anda, Managementul riscului la patrunderea pe piata internationala, Editura Victor, Bucuresti, 2009, pag.251*

F is a more complex factor and can have values between 0.1 (absolutely incompatible ) and 100 (total compatibility).

The ratio $\frac{1+RI_T}{1+RI_O}$ reflects the monetary risks; if $RI_O > RI_T$, it is a signal that the country of origin is weaker than that of the target-market. The reasonable limits of the scale are 0 and 2; over this value, the inflation of the target-market is soaring, the market entry risk being very high.

The ratio $\frac{1+RCE_T}{1+RCE_O}$ reflects the risk of entering a market with a different rhythm of growth than that of the country from where the investment comes. The reasonable limits of the scale are o and 2; over this value, the inflation of the target-market is plummeting, the market-entry risk being too high.

The ratio $\frac{V\,company\,x}{CS\,company\,x}$ , varies between 0 şi 100.

If $V_{company\,X} > CS_{company\,X}$, the company is rich in assets, which exceed the scriptural value of the capital (most common situation), while if $V_{company\,X} < CS_{company\,X}$, the company is almost bankrupt.

The indicator I can take positive values (from 0 to +∞).
For a nuanced analysis, one can apply apply the logarithm over I and the result is

$I^* = lg\,I = lg\,(N \times \frac{1+RI_T}{1+RI_O} \times \frac{1+RCE_T}{1+RCE_O}) + lg\,(F \times \frac{V\,company\,x}{CS\,company\,x})$.

The first term characterizes the degree of risk of the target region/country, while the second characterizes the microeconomic risk.

If $I^* < 0$ and the country risk is more than 6, the factors which characterize the external environment being in normal limits, than the company envisaged for association is either less evaluated, or almost bankrupt, vulnerable, able to be taken over very easily and changed radically, in terms of items produced.

If $I^* > 5$, under the same circumstances, than the company envisaged for association has a very good financial situation.

Therefore, five intervals of values for the $I^*$ indicator were settled. The extremes of this grid are:
$I^* < 0$, in this case, the optimal strategy is the direct greenfield investment
$I^* > 5$, in this case, the optimal strategy is the export, as it can be seen in the following table:

**Table 1-The evaluation of the microeconomic environment analysed in rapport with the values on the grid of I$^*$**

| The values of I$^*$ =lg(I) | The evaluation of the microeconomic environment | The optimal entry strategy |
|---|---|---|
| I$^*$ < 0 | The microeconomic environment likely to be entirely taken over | Direct greenfield investment |
| 0 < I$^*$ < 1,6 | The microeconomic environment likely to be entirely taken over by a buy of the majority of stocks and joining the management team | Acquisition |
| 1,6 < I$^*$ < 2 | The microeconomic environment likely to be taken over at a equal rate to that of the partner | Mergers, acquisitions |
| 2 < I$^*$ < 5 | The microeconomic environment favourable for economic cooperation | Licensing, franchising, strategic alliances, management contract |
| I$^*$ > 5 | The microeconomic environment hard to be approached through a partnership but favourable for trading operations | Export |

Of course, there are many other tools used in the investment process, like methods based on the experience and subjective assessment of the analyst/team/consultant, which apply the cost-profit analysis and which have set as an objective the examination of different elements for the risk calculus[2].

### 3. The case of Electroputere SA-Al Arrab

In order to illustrate the above described analysis tool, we have chosen the case of Electroputere S.A., the old Romanian producer of railway equipment, for which we calculated the time evolution of the indicator I *, then we made predictions for further 4 years, by using the extrapolation technique.

In this regard, by using the database provided by Financial Investment Services Company SSIF Intercapital SA, we have calculated daily quotations and the volume of shares transacted by Electroputere SA on the Bucharest Stock Exchange.

Electroputere Craiova was founded on September 1, 1949, producing equipment for the energy sector and for the rail transport sector. In 1990, Electroputere was divided into 7 companies and in 1994 they met in the current holding SC Electroputere SA. The company still produces motors, electric apparatus, rotating electrical machines, power transformers, railways and urban vehicles. In 2007, the local sales were 49%, while the overseas sales were 41%. Unfortunately, in time, the external market, has decreased slowly, because the management politics have been not set enough on conquering new areas; for instance, the company did not participated to many foreign auctions, which could have helped it to grow.

Electroputere Craiova was formerly owned by the state assets recovery body (AVAS) and in 2007 the company was privatized, after a public bid. Before the privatization, the shareholders structure was the following:

---

[2] *Gheorghiu Anca, Econofizică investiţională, Editura Victor, Bucureşti, 2007, pag.203*

| Shareholder | Shares | Percentage |
|---|---|---|
| AVAS | 78.009.632 | 62,83% |
| ALȚI ACȚIONARI | 34.536.922 | 28,38% |
| CLIENTI NEREZIDENȚI-CITIBANK NOMINEE A/C | 11.621.400 | 8,80% |
| **Total** | **124.167.954** | **100%** |

*Fig. 1- The shareholders structure at Electroputere SA at Oct. 31, 2007*
*Source: Depozitarul Central (www.bvb.ro)*

Al-Arrab Construction Company Ltd , a firm from Saudi Arabia took over on November 6, 2007, 62.82% of the equity of the company Electroputere. It was a 120 million euro deal. According to AVAS, Al-Arrab Contracting Company Ltd Saudi Arabia has made an investment program amounting to 23,084,000 euros, of which EUR 3,084,000 EUR 20,000,000 investment environment and investment for development and refurbishment.

Al-Arrab Construction Company is a company located in Saudi Arabia which aims at reaching the highest possible percentage of the construction industry professional active in the Middle East and North Africa (MENA) building, construction machinery, road, and power generation markets. Al-Arrab consider the investment as a step for entering the very important and strategic market of Romania and intends to transform Electroputere into a global producer of energy and transport. The group of Al-Arrab is composed of 30 firms valuing 40 billion (27 billion euros). Al-Arrab had, in 2006, a turnover of 133.85 million dollars (90 million euros) and a net profit of 13.65 million dollars (9.2 million euros).

The evolution of the quotation of Electroputere over the capital market, the indicator I and its decimal logarithm (I *) increased before the privatization. Just before the takeover, it recorded a dramatic boost, probably due to a positive perception of the other investors. In other words, although the company has had not yet being profitable, the market anticipated good future prospective.

Therefore, it is very interesting to analyze the development of indicators I and I * before the takeover of Electroputere by Al-Arrab Contracting Company, namely at the date of June 1, 2007 (a randomly chosen date, with no special meaning) and November 6, 2007, the date when the company has been taken over.

We took into account the following factors:
- N-the rating which expresses the level of risk for the target country (Romania), in value of 7, because in 2007, Moody's rating for Romania was Baa3.
- F-the score given for the degree of cultural/organizational compatibility, in value of 10 (compatible) because, although there were important cultural differences between our country and Saudi Arabia, in terms of business international experience and management style, there are approaches and therefore compatible.
- $RI_T$ -inflation rate in the target-country, Romania, 6.6% in 2007.
- $RI_O$ -inflation rate, in Saudi Arabia, 4.1% in 2007 *(Source: CIA Worldfactbooks[3])*
- $RCE_T$ - growth rate of the target country, Romania, 6% in 2007

---

[3] *https://www.cia.gov/library/publications/the-world-factbook/geos/sa.html*

- RCE₀ - growth rate in the country of origin, Saudi Arabia, 4% in 2007 *(Source: CIA Worldfactbooks)*
- CS $_{Electroputere}$, the common stock of Electroputere S.A., amount: 12.416.795,40 lei
- V $_{Electroputere}$, the economic value of the enterprise.

| Date | Market capitalisation | Number of shares | V $_{Electroputere}$ |
|---|---:|---:|---:|
| 01.06.2007 | 0,256 | 124.167.954 | 31.786.996 lei |
| 06.11.2007 | 1,22 | 124.167.954 | 151.484.904 lei |

*Fig. 2- The economic value of Electroputere S.A. in 2007*
*Source: The authors*

The value of the indicator at the two dates chosen for analyze is the following:

June 1, 2007:

$$I = N \times F \times \frac{1+RI_T}{1+RI_O} \times \frac{1+RCE_T}{1+RCE_O} \times \frac{V_{firma\ EPT}}{CS_{firma\ EPT}} = 7 \times 10 \times \frac{1+0,066}{1+0,041} \times \frac{1+0,06}{1+0,04} \times \frac{31.786.996\ lei}{12.416.795\ lei} = 4,382$$

I* = lg(I) = lg 4,382 = 0,641672373

November, 6, 2007:

$$I = N \times F \times \frac{1+RI_T}{1+RI_O} \times \frac{1+RCE_T}{1+RCE_O} \times \frac{V_{firma\ EPT}}{CS_{firma\ EPT}} = 7 \times 10 \times \frac{1+0,066}{1+0,041} \times \frac{1+0,06}{1+0,04} \times \frac{151.484.904\ lei}{12.416.795\ lei} = 20,8895$$

I* = lg(I) = lg 20,8895 = 1.31992804

  We notice that on both dates, namely on June 1, 2007 and the November 6, 2007, the value of lg (I) fluctuates within the interval [0-1,6], which convey that in 2007, Electroputere SA was a company predisposed to be taken over. Therefore, the indicator confirms that the market entry strategy chosen by Al-Arrab Construction Company, i.e. acquiring a majority stake in Electroputere SA, was a correct one.

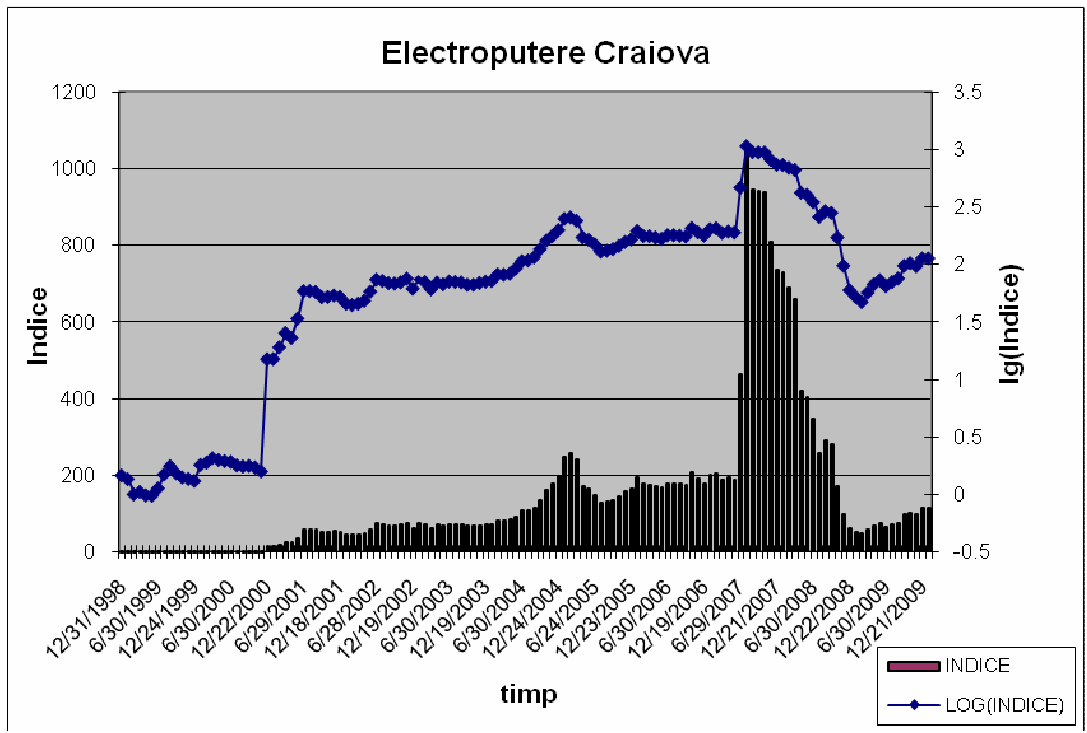

*Fig.3-The evolution of I and I\*for Electroputere S.A.*
*Source: Data from Intercapital Start www.intercapital.ro and Tradeville www.tradeville.eu; data processing was made by the authors*

Fact is that, after the takeover, the market capitalization of Electroputere SA declined, as happened with many other companies, not only in Romania, but also worldwide. It was a general trend in 2007, when public perception was a negative one and this affected the market value of the large majority of corporations, due to the U.S. subprime crisis, which had influenced global macroeconomic indicators. Subprime lending in finance means making loans that are in the riskiest category of consumer loans and are typically sold in a market from prime loans. In the early 2000's, banks credited many people with loans based on uncertain incomes; they were people with doubtful creditworthiness.

The crisis began to affect the financial sector in February 2007, when HSBC, one of the world's largest bank, wrote down its holdings of subprime-related MBS by $10.5 billion, the first major subprime related loss to be reported. The main effect of the crisis was a much diminished access to cash, therefore the crisis has affected many businesses, especially investment banks, such as Lehman Brothers or brokerage corporations such as Merrill Lynch.

For 2008 and 2009, we have calculated the indicators I and I * by taking into account some changes of the input data:
a) because of the world economic downturn, Romania has been severely affected economically in the past 15 months and some rating agencies have downgraded the country to "junk" level for investment; that is the reason for which we have considered a level of 5 for N (the country-risk rating of the target-market)
b) the inflation level for the target-market (Romania- $RI_T$ ) has grown in 2008 and 2009 from 6% to 10-12%
c) we considered that the company may be interesting for acquisition by a company situated in EU, therefore we have taken into account a rate of economic growth for the country of origin ( $RCE_O$ ) of 1%, while for Romania, we have considered a rate of economic growth for the target-country ( $RCE_T$ ) of -7%.
d) we considered F = 10 (complete cultural/organizational compatibility.

The outcome demonstrates that during the crisis, I and I * have both sharply fallen, with a negative record in the summer of 2009. No wonder, that the company announced in August 2009 a plan to dismiss 500 employees, following the company's restructuring plan. Then, it was a very favourable period for acquisition. Still, because of the acute lack of liquidities and of the country risk downgrading, no foreign company had shown any sign of interest for Electroputere. However, since the indicators are still low, the company remains very attractive for investors in the first quarter of 2010.

The evolution of the market value of Electroputere S.A shows also a significant drop after September 2009, with timid signs of recovery in the last months of 2009.

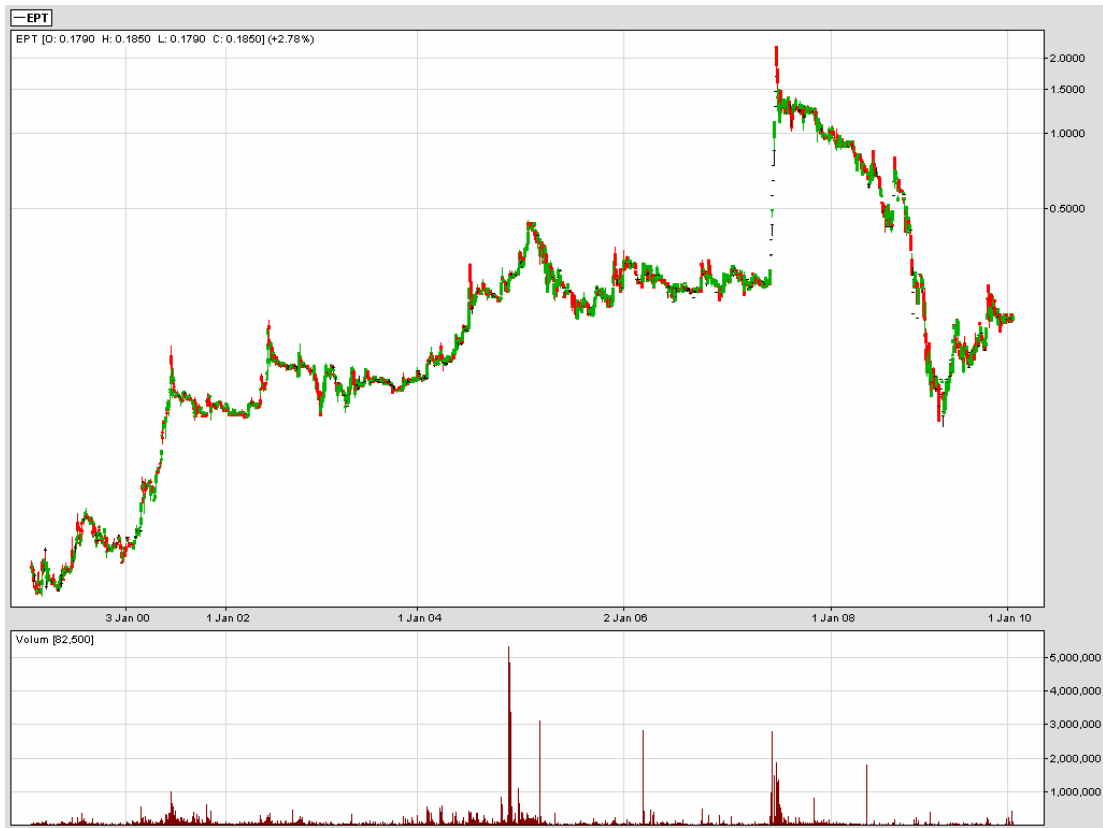

*Fig.4- Evolution of the market value of Electroputere S.A. (EPT)*
*Source: Intercapital Start www.intercapital.ro and Tradeville www.tradeville.eu; data processing was made by the authors*

## 4. Later developments and conclusions

The later developments showed that, despite a good evolution in the first months of 2007, the effects of the global crisis have been felt in Romania as well and have had a negative influence over the evolution of Electroputere stocks (see fig.5). The most dramatic fall was in the first quarter of 2009. Afterwards, one can notice a rising trend of the price evolution, which will probably continue in the following months.

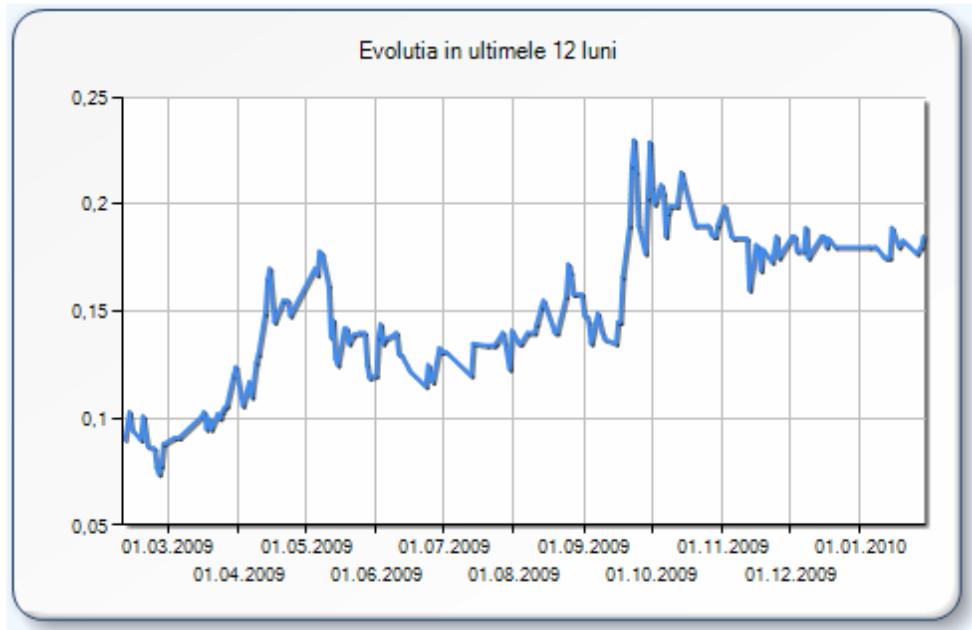

*Fig.5- Price evolution of Electroputere SA stocks in 2009-2010*
*Source: www.bvb.ro*

Al Arrab Company Ltd offered 2.3 million euros in 2008 for 62.82 % of the shares, but ended up by owing 86.28 percent of the company after the last share capital increase (December 2008). The up-to-date number of shares of Al Arrab at December 31, 2009 is 291.284.640 (86,28%) from the total capital.

Electroputere Craiova announced losses of 545,300 lei in the first quarter of 2009, down 13-fold compared the same period last year. Its business advanced from 28.3 million lei to 63.18 million lei in the mentioned period.

As a conclusion, assessing the risk the prospective of success or failure for a company trying to enter a new market is a must. There are many tools, models, algorithms or software for risk management; among them is the indicator presented in the present paper. The case of Electroputere S.A. is relevant; it shows how the analyst should proceed in order to find and propose to the management team the best strategy for entering a new market. Definitely, no model or method can offer a 100% guarantee that the project will be a success. Unpredictable events, like the subprime crisis which started in 2007 and evolved as an international financial crisis, can shatter the plans of a company to enlarge the global market share and heighten the profits.

Well-capitalized companies, with robust financial positions and with liquidities to invest, have now a unique chance to make strategic acquisitions and to buy previously unaffordable assets. Of course, it takes audacity to act in a risky climate, but as more and more targets become available, taking advantage of this occasion will help daring companies to have success in the years of boost after the crisis.

A recent study[4] conducted for Deloitte Touche Tohmatsu by Forbes Insights reveals that economic optimism reached its highest level in the past 12 months among surveyed executives only in December 2009. The analysis discloses that in the last month of the year 2009, more than one-third of the executives surveyed (35%) believe the worst of the economic crisis is behind us-the highest level of economic confidence since the survey began in January 2009 (Figure 6). The data comes from a survey of 335 senior executives and talent managers from large companies (annual sales $500+ million), across a range of many industries and the three major economic regions, the Americas, Asia Pacific (APAC), and Europe, the Middle East, and Africa (EMEA).

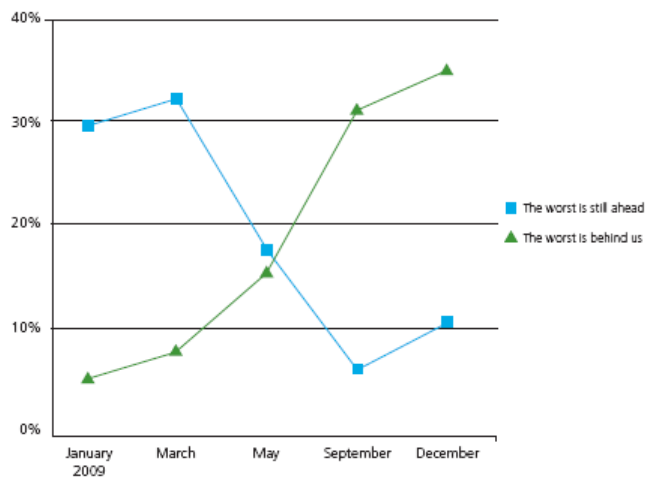

*Figure 6-Executive outlook on the economy (December 2009)*
*Source: Deloitte Development LLC*

Indeed, the forecast of IMF[5] suggests the same positive trend will follow the major economic indexes. The world trade volume is expected to rise in 2010 at 5.8 per cent in 2009 compared with a dramatic -12.3 per cent in 2009. The oil prices is predicted to get much higher in 2010 (22.6 per cent in 2010 compared to a sharp drop of -36.1 in 2009). Projections are as well good for 2011.

---

[4] *Deloitte Development LLC-Managing talent in a turbulent economy – January 2010*
[5] *International Monetary Fund- Overview of the World Economic Outlook Projections, 2010, http://www.imf.org/external/pubs/ft/weo/2009/update/01/pdf/table1.pdf*